\documentclass[aps,prx,a4paper,twocolumn,longbibliography,floatfix,superscriptaddress]{revtex4-2}

\usepackage{amssymb,amsmath,amsfonts}
\usepackage{ dsfont }
\usepackage{comment}
\usepackage{amsxtra}
\usepackage{accents}
\usepackage{graphicx}
\usepackage[dvipsnames]{xcolor}
\usepackage[bookmarksopen,colorlinks,linkcolor=blue,citecolor=olive,urlcolor=red]{hyperref}
\usepackage{bm}
\usepackage{amssymb}
\usepackage{amsmath}
\usepackage{amsfonts}
\usepackage{graphicx}
\usepackage{bm}
\usepackage[dvipsnames]{xcolor}
\usepackage{natbib}
\usepackage{hyperref}
\usepackage{comment}
\usepackage{moresize}

\def\vec k{\mathbf{k}}

\newcommand\smvee{\raise0.8ex\hbox{$\scriptscriptstyle\vee$}}

\def\vec k{\mathbf{k}}

\newcommand{\eps}{\varepsilon}

\newcommand{\addLL}[1]{\textcolor{red}{#1}}
\newcommand{\addQ}[1]{\textcolor{magenta}{#1}}
\newcommand{\la}{\left\langle}
\newcommand{\ra}{\right\rangle}

\renewcommand{\vec}[1]{{\boldsymbol #1}}
\newcommand{\lp}{\left(}
\newcommand{\rp}{\right)}

\def\be{\begin{equation}}
\def\ee{\end{equation}}
\begin{document}
\title{Berry-Flux-Controlled Cascade of Chiral Superconducting States} 
\author{Daniil Karuzin}\affiliation{Moscow Institute of Physics and Technology, Dolgoprudny 141701, Russia}
\altaffiliation[also:\ ]{Landau Institute for Theoretical Physics, Moscow, Russia}

\author{Zhiyu Dong}\affiliation{Department of Physics, California Institute of Technology, Pasadena CA 91125}

\author{Leonid Levitov}\affiliation{Department of Physics, Massachusetts Institute of Technology, Cambridge MA02139, USA} 
\altaffiliation[email:\ ]{levitov@mit.edu}

\begin{abstract}
Motivated by recent interest in chiral superconductivity in narrow bands, we develop a general framework to clarify how band topology and quantum geometry affect superconducting pairing and connect to the two-body problem.
Berry curvature does not merely favor a chiral pairing channel; it produces a sequence of chiral pairing instabilities indexed by angular momentum, controlled by the Berry flux through the Fermi sea, with a Little-Parks-like periodicity in momentum space.
%
%
We show that Berry curvature converts a nonchiral attractive interaction into a geometrically frustrated Cooper problem in momentum space. The relevant control parameter is the Berry-curvature flux enclosed by the Fermi sea, $\Phi = b k_F^2$, which acts as an effective Aharonov–Bohm flux for the order parameter defined on the Fermi surface. As $\Phi$ is tuned, the leading pairing instability switches between odd angular-momentum channels $m=1,3,5…$, producing a cascade of first-order transitions and Little–Parks-like oscillations of $T_c$.
\end{abstract}

\maketitle

Since its inception, graphene has been widely explored as a platform for exotic
superconductivity~\cite{
Uchoa2007,BlackSchaffer2007,Honerkamp2008,Pathak2010,
Kiesel2012,Uchoa2013,Nandkishore2012}. 
This idea gained firm support with the discovery of superconductivity in twisted bilayer graphene~\cite{Cao2018MAGSC,
      Chen2019TLG,
      Lu2019MAG,
      Stepanov2020MAG,
      Yankowitz2019TBG,
      Zhou2022BBG}, 
      followed by correlated states breaking time-reversal symmetry in rhombohedral graphene~\cite{Choi2025SCQAH,
      Guo2025ThickRMG,
      Han2023Multiferro,
      Holleis2025BBG,
      Kumar2025DualSurface,
      Li2024BBG,
      Patterson2025SOCrtg,
      Xie2025BFieldRMG,
      Yang2025SOCrmg,
      Zhou2021RTGSC,
      Zhou2022BBG}.
Most recently, direct signatures of chiral (time-reversal-broken) superconductivity have been reported
in rhombohedral graphene tetralayer and pentalayer~\cite{Han2025ChiralRMG},
followed by a flurry of theoretical work probing various aspects of
chiral superconductivity, with an emphasis on the effects of
Berry phase and unconventional pairing mechanisms
\cite{WangEtAl2021,
Murshed2025,ChouEtAl2025,ParraMartinez2025,Yang2024,Qin2024,
Shavit2025,Geier2024,Jahin2024,Guerci2025,Dong2024, 
MatsyshynEtAl2024,MayMannEtAl2025,  GhazaryanEtAl2023,ChouEtAl2025,BostromEtAl2024,GaggioliEtAl2025,
PaolettiEtAl2025,MayMannEtAl2025,Jahin2026KL}.

\begin{figure}[tb]
    \centering
    \includegraphics[width=0.99\linewidth]{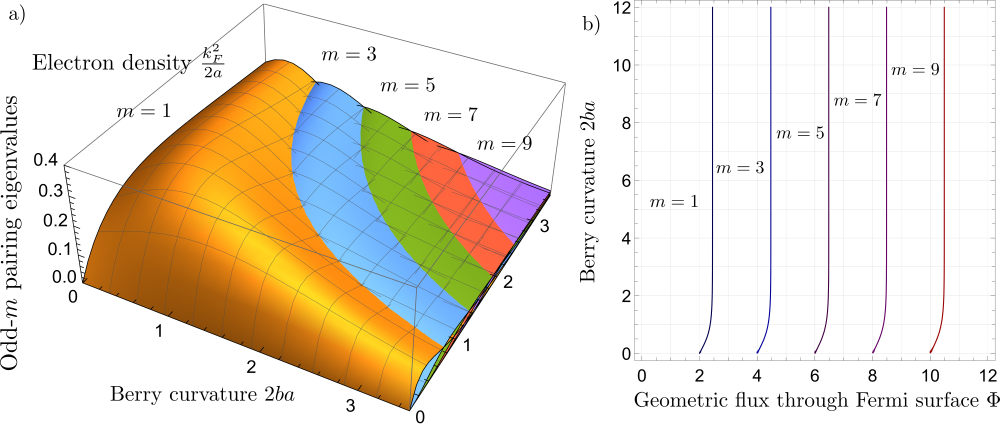} 
   \caption{Berry-flux-controlled cascade of chiral superconducting states.
   (a) Leading odd-$m$ pairing eigenvalues $g_m$ of the linearized gap equation as functions of Berry curvature $b$ and carrier density $k_F^2/4\pi$. Increasing the Berry flux $\Phi=bk_F^2$ drives first-order transitions between chiral states with $m=1,3,5,…$.
(b) Phase boundaries $g_m=g_{m+2}$ plotted versus $\Phi$, showing the approximate Little–Parks-like periodicity with periodicity $\Delta\Phi=2$. The boundaries are well approximated by constant-$\Phi$ lines, a family of hyperbolas in a) and vertical lines in b); in the large-$a$ limit they approach $\Phi = \sqrt{(m+1)(m+2)}$, while in the small-$a$  limit they approach the integer-flux condition discussed near Eq.\eqref{eq:LP_oscillations}.
    }
\vspace{-4mm}
    \label{fig1}
\end{figure}

\begin{figure}[tb]
    \centering
    \includegraphics[width=0.8\linewidth]{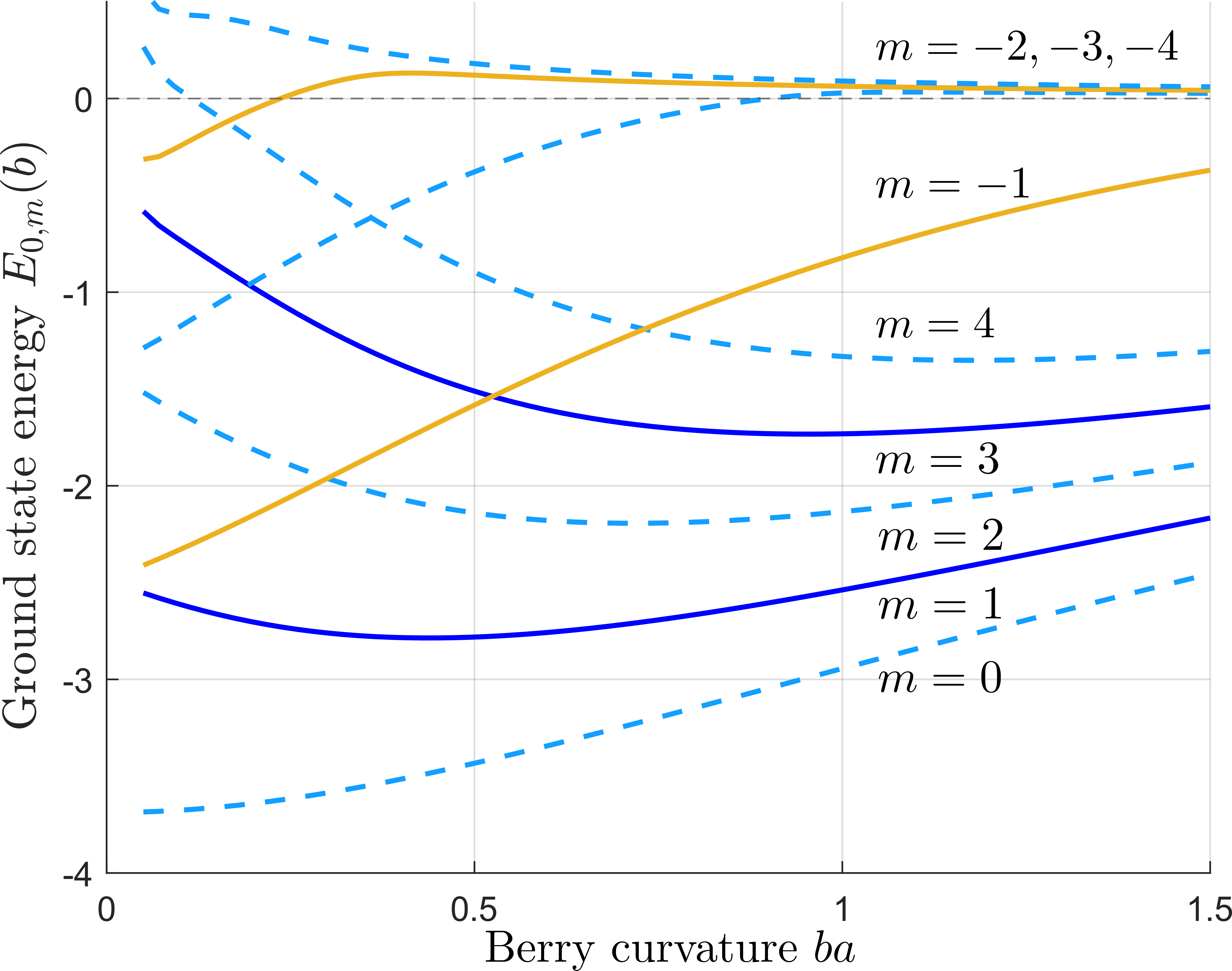} 
    \caption{
    Ground-state energies of the two-body problem, Eq.~\eqref{eq:H12}, 
as functions of $b$ for different angular-momentum sectors $m$. 
Solid lines show odd $m$ and dashed lines show even $m$ for representative parameters. 
For fermions, even-$m$ and odd-$m$ channels correspond to spin-singlet and spin-triplet 
states, respectively. States with $m$ and $-m$ are degenerate at $b=0$ but become 
nondegenerate at finite $b$. Finite Berry curvature favors $m>0$ states, whose energies 
decrease (become more negative) as $b$ increases. The apparent bound states at $E>0$ 
are a finite-size effect. Notably, the ground states with different $m$ do not alternate 
as $b$ varies, indicating that the alternation of pairing channels in Fig.~\ref{fig1} 
instead arises from geometric flux and the Fermi-sea area commensurability (see 
Eq.~\eqref{eq:GL_model} and discussion).
    }
\vspace{-4mm}
    \label{fig2_spectrum}
\end{figure}

Here, our goal is to highlight some unique aspects of the 
quantum-geometric pairing problem in bands with Berry curvature 
arising due to universal geometric phases acquired by Cooper pairs during scattering. We begin with the two-particle bound-state problem (Fig.~\ref{fig2_spectrum}) and then extend it to the pairing problem (Fig.~\ref{fig1}). 

The central result is that Berry curvature does more than select a chiral $p+ip$ state. Because the superconducting order parameter lives on the Fermi surface, the Berry flux enclosed by that surface acts as an effective momentum-space flux. The preferred winding number $m$ is therefore selected by a commensurability condition between $m$ and $\Phi$, producing a cascade of chiral phases rather than a single chiral instability.

We develop a microscopic framework in which Berry curvature enters Cooper scattering through universal geometric phases. In the constant-curvature model used below, the phase accumulated in scattering $(\vec k,-\vec k)\to (\vec k',-\vec k')$
is proportional to the momentum-space area $(\vec k'\times \vec k)_z$. This converts an otherwise nonchiral attraction into a chiral pairing interaction and produces the phase diagram shown in Fig.~\ref{fig1}.

First-order transitions between these phases give rise to a ``quantum-geometric'' Little-Parks-like effect, manifested as oscillations of the critical temperature $T_{\rm c}$ as a function of the Berry-curvature flux enclosed by the Fermi sea 
(times two to account for pairing effects), 
\begin{equation}
\label{eq:Phi_FS}
\Phi = 2\int_{FS} \frac{d^2 k}{2 \pi} \Omega(\vec k)  \equiv b k_F^2
    ,
    \end{equation}
    with $\Omega(\vec k)$ the band Berry curvature. 
For a chiral Dirac band 
$\Omega(\vec k)$ is generally momentum dependent, whereas
%
the constant-curvature model used below
reduces this expression to $\Phi=bk_F^2$. We therefore expect the qualitative
mechanism discussed below---the geometric phase in Cooper scattering and the
commensurability between Fermi-surface winding and enclosed Berry flux---to
persist for momentum-dependent Berry curvature, although the detailed locations
of the phase boundaries need not remain identical to those of the constant-$b$
model.

It is instructive to draw a comparison with conventional type-II superconductors, where vortices induced by magnetic fields do not fundamentally alter the microscopic structure of the paired state; their main effect is to generate vortex phases and impact the long-range order of the system~\cite{Tinkham1996,Blatter1994,Brandt1995,CrabtreeNelson1997,Kwok2016}. 
By contrast with the conventional case, Berry curvature directly influences pairing type and symmetry on microscales. Berry-curvature-induced pseudomagnetic fields also drive transitions between distinct topological superconducting phases, in which the order-parameter phase winds in momentum space and the superconducting state acquires nontrivial topology~\cite{ReadGreen2000,KallinBerlinsky2016,SatoAndo2017,MayMannEtAl2025}. Such phases are expected to exhibit strong orbital magnetism, host exotic (Majorana) quasiparticles and chiral edge excitations, and display anomalous thermal transport and related signatures~\cite{ReadGreen2000,NayakEtAl2008,KallinBerlinsky2016,SatoAndo2017}. The goal of this work is to elucidate how Berry curvature reshapes the phase diagram of superconductors, promoting chiral, time-reversal-breaking states even in the absence of any applied magnetic field.

To understand the key aspects of the chiral pairing problem, 
we first consider the two-body problem 
\begin{equation}\label{eq:H12}
    H_{12} = \frac{\vec k_1^2}{2m}+\frac{\vec k_2^2}{2m} + V(\vec R_1-\vec R_2)
,\quad \vec R_j=\vec r_j+\vec a(\vec k_j)
,
\end{equation}
where
$
    \vec a(\vec k) = i\langle u_{n\vec k}|\nabla_{\vec k}u_{n\vec k}\rangle
$
is the Berry connection of band $n$, and we set $\hbar=1$.
The operators $\vec R_j$ denote covariant particle coordinates~\cite{Sundaram1999,Xiao2010,Panati2003,TeufelBook}.
This Hamiltonian can be viewed as the single-band projection of an underlying
multiband Dirac-type Hamiltonian, where the covariant-derivative structure
$
    \vec R_i = i\partial_{\vec k_i} + \vec a(\vec k_i)
$
emerges within the adiabatic approximation~\cite{Sundaram1999,Xiao2010,Panati2003,TeufelBook}.

The covariant guiding-center form in Eq.~\eqref{eq:H12} should be understood as the
leading quantum-geometric limit of the 
single-band projected
interaction. In the standard formulation, the interaction projected to an
isolated Bloch band is dressed with coherence factors, or form factors,
\be\label{eq:Fkk'}
    F_{\vec k,\vec k'}=\langle u_{\vec k}|u_{\vec k'}\rangle
    ,
\ee
%
%
which dress the Cooper-channel vertex as 
\[
    V_{\vec k,\vec k'}
    =
    V(\vec k'-\vec k)F_{\vec k,\vec k'}F_{-\vec k,-\vec k'}
    -
    V(\vec k'+\vec k)F_{\vec k,-\vec k'}F_{-\vec k,\vec k'} .
\]
This direct-minus-exchange structure will be used below in the Cooper channel; the associated two-body Hamiltonian 
clarifies the geometric phase content. 
For small momentum transfer $\vec q$, the overlap has the expansion
\be\label{eq:Fkk'_phase_repres}
    F_{\vec k,\vec k+\vec q}
    =
    \langle u_{\vec k}|u_{\vec k+\vec q}\rangle
    \simeq
    \exp\!\left[i\vec q\cdot \vec a(\vec k)\right]
    + O(q^2),
\ee
where $\vec a(\vec k)=i\langle u_{\vec k}|\nabla_{\vec k}u_{\vec k}\rangle$
is the Berry connection. These Berry-connection phases are precisely
reproduced by replacing the canonical coordinate $\vec r$ by the covariant
wave-packet coordinate
\be
    \vec R=\vec r+\vec a(\vec k).
\ee
Thus Eq.~\eqref{eq:H12} is not a separate phenomenological ansatz, but rather
a minimal geometric representation of the projected-band interaction. It
isolates the Berry-phase contribution to Cooper-pair scattering; higher-gradient
corrections contain additional quantum-geometric data, including the quantum
metric.


We stress that such quantum-geometry structure of the interaction is totally general, applicable to any band dispersion and any Berry connection distribution in $k$ space, applicable both for the ordinary case when time reversal symmetry (TRS) is unbroken and when it is broken by valley polarization (this paper). TRS imposes the relation 
$
    \vec a_K(\vec k) = -\vec a_{K'}(-\vec k)
$ 
between the Berry connections in the $K$ and $K'$ valleys. As a first step toward
analyzing superconductivity in valley-polarized graphene (discussed below), we
now consider two interacting electrons in the same valley.


We focus on the simplest case of a cylindrically symmetric interaction $V(\vec R_1 - \vec R_2)$, for which individual angular-momentum channels can be solved exactly.
Using translation symmetry, we work in momentum space, $\vec{p}_j \to \hbar \vec{k}_j$, $\vec{r}_j \to i \partial_{\vec{k}_j}$, $j=1,2$. We 
go to 
the center-of-mass frame and use 
relative coordinates and momenta:
\begin{equation}
    \vec{r} = \vec{r}_1-\vec{r}_2, \quad \vec{k} = (\vec{k}_1-\vec{k}_2)/2.
\end{equation}
The two-particle Hamiltonian in the center of mass frame $\vec k_1+\vec k_2=0$ takes the form of a one-particle Hamiltonian for a particle with a reduced mass $\mu=m/2$:
\begin{equation}
\label{eq:H12_CM}
    H_{12} = \frac{\hbar^2\vec k^2}{2\mu} + V(\vec R_1-\vec R_2),\quad
  \vec R_j=i \partial_{\vec k_j}+\vec{a}(\vec{k}_j). 
\end{equation}
Here, for illustration, we consider a Gaussian model
\begin{equation}
\label{eq:V12}
    V(\vec R_1- \vec R_2) = \lambda e^{-a(\vec R_1-\vec R_2)^2} 
    ,\quad \lambda<0, 
\end{equation}
and choose two-particle Berry connection $
\vec a_{2e} (\vec k)\equiv \vec a(\vec k)-\vec a(-\vec k)$. 
For a uniform field 
\be
\vec a_{2e} (\vec k)=
bk\,\bm e_\varphi =  b(-k_y,k_x)
.
\ee
In this case, the chiral two-particle problem can be solved in a closed form. To that end, we consider the operator
\begin{equation}
 \vec R^2 
  = \left( i  \partial_{\bm k} + b k\,\bm e_\varphi \right)^2.
  \label{eq:R-operator}
\end{equation}
This quantity is nothing but the Hamiltonian of a non-relativistic particle in an external magnetic field, with eigenvalues $\rho_{n,m} = 2 b  \left(
   2 n + |m| + m + 1
  \right)$. The spectrum of the problem $\left. V(\vec R)|\psi\ra=\left. \epsilon|\psi\ra$ is therefore
\begin{equation}
\epsilon_{n,m}=\lambda e^{-a \rho_{n,m}},\quad
n\ge 0, \quad m=0,\pm1,\pm2...,
\label{eq:Enm}
\end{equation}
and the corresponding eigenfunctions in momentum space can be written as
\begin{align}
\label{eigen}
  \psi_{n,m}(k,\varphi)=
  c e^{-i m \varphi} e^{-\frac{b k^2}{2}}
    k^{|m|}
    {}_1\! F_1\!\left(-n,\,|m| + 1,\,b k^2\right),
    \nonumber
\end{align}
where $ {}_1\! F_1$ denotes a confluent hypergeometric function, and $c=b^{\frac{1+|m|}{2}}
    \sqrt{\frac{(|m| + n)!}{ \pi n! \, |m|!^2}}$
    is normalization factor.

The Hamiltonian in Eq.~\eqref{eq:H12_CM}, owing to its cylindrical symmetry, decomposes into independent blocks labeled by angular momentum $m$. In each 
$m$-sector, it can be represented in the circular basis $\psi_{n,m}$ as a real symmetric tridiagonal matrix (here $n,n'\ge 0$ and $m$ is arbitrary):
%
%
%
\begin{align}
& \la n'|H^{(m)}_{12}|n\ra=
g_n\delta_{n'n}+\tilde g_{n}\delta_{n',n-1}
+\tilde g_{n'}\delta_{n',n+1}
\\ &
g_n=\lambda e^{-a \rho_{n,m}}+\frac{
2n + |m| + 1}{2\mu b}
,\quad
\tilde g_{n}=-\frac{\sqrt{n(n+|m|)}}{2\mu b}
.
\nonumber
\end{align}
%
Diagonalizing this Hamiltonian yields the ground states in each 
$m$-sector as functions of 
$b$, as shown in Fig.~\ref{fig2_spectrum}. Notably, the states with angular momenta 
$m$ and 
$-m$ split into pairs of nondegenerate levels that (initially) disperse linearly with 
$b$, one moving up and the other down in energy, whereas the $m=0$ state remains the ground state for all $b$. As $b$ increases, the energies of the positive-$m$ states become more negative, while the energies of the negative-$m$ states increase and eventually merge with the positive-energy continuum. In the continuum, $H_{12}$ 
describes a chiral scattering problem that gives rise to chiral pairing and superconductivity.

The tendency of Berry curvature to favor nonzero-chirality states is key to understanding chiral superconductivity. However, as a comparison between Figs.~\ref{fig1} and \ref{fig2_spectrum} shows, the leading-$m$ superconducting pairing channel cannot be inferred directly from the properties of the two-body problem, Eq.\eqref{eq:H12}. The reason is that new effects arise in the presence of a Fermi surface, where 
pairing actually develops. In particular, 
superconductivity becomes sensitive to a commensurability condition between the
Fermi sea area and enclosed Berry flux. 
As we will see, this commensurability drives 
quantum-geometric Little-Parks oscillations 
manifested in the succession of superconducting states with different $m$.

The two-body problem is therefore diagnostic rather than predictive. It reveals the microscopic origin of chiral scattering from Berry curvature, but not which superconducting angular-momentum channel wins. That selection requires a Fermi surface, since the relevant variable is the Berry flux enclosed by the occupied states.

We now proceed to analyze superconductivity in a 2D Fermi sea 
with a quantum-geometric pairing interaction
\be
H=\sum_{\vec k} \xi(\vec k) c_{\vec k}^\dagger c_{\vec k}+\sum_{\vec k, \vec k'}  \Gamma_{\vec k, \vec k'} 
c_{\vec k'}^\dagger c_{-\vec k'}^\dagger  c_{-\vec k} c_{\vec k}
\ee
$\xi(\vec k)=\frac{\vec k^2}{2m}-\mu$. 
We focus on  pair-scattering processes
$(\vec k,-\vec k)\to(\vec k',-\vec k')$ that drive the pairing instability.
For simplicity, we ignore valley and spin degrees of freedom, taking fermions as spinless. This framework is relevant for rhombohedral graphene superconductors in which electrons are spin- and valley-polarized. Spin-triplet pairing mandates odd-parity angular momentum $m$, limiting possible $m$ to odd values $m=1, 3, 5,...$. 
The model can be readily extended to include spin.



We now apply the projected interaction of Eqs. \eqref{eq:Fkk'}--\eqref{eq:Fkk'_phase_repres} to the Cooper channel. The resulting pairing vertex has the antisymmetrized form
\be \label{eq:Vkk'_preview}
    \Gamma_{\vec k,\vec k'}
    = \Gamma^{\rm dir}_{\vec k,\vec k'}
      - \Gamma^{\rm exch}_{\vec k,-\vec k'} 
      ,
\ee
where the geometric phases contained in the two terms are inherited from the Berry-connection factors discussed above. For the constant-curvature model, the direct term contains the phase $e^{i b e_{\vec{z}} (\vec k' \times \vec k)}$, while the exchange term contains the opposite phase. After projection onto angular momentum $m$, antisymmetrization gives cancellation in even-$m$ channels and constructive addition in odd-$m$ channels. Thus, for the spin-polarized triplet problem considered here, only odd $m=1,3,5,…$ remain and these couplings are doubled.
This 
yields 
a chiral paired state described by the Hamiltonian
\be
H=\sum_{\vec k}\xi_{\vec k} c^\dagger_{\vec k} c_{\vec k}+\sum_{\vec k}\Delta^*(\vec k) c_{-\vec k}c_{\vec k}+\Delta(\vec k) c^\dagger_{\vec k}c^\dagger_{-\vec k},
\ee
where complex-valued $\Delta(\vec k)$ obeys the gap equation 
\begin{equation}\label{eq:gap_equation}
    \Delta(\bm k) =  -\sum_{k'} \Gamma_{\vec k, \vec k'}  \frac{\Delta(\bm k')}{2 E_{\bm k'}} \tanh\bigg(\frac{ E_{\bm k'}}{2T}\bigg), 
\end{equation}
$E_{\bm k'} = \sqrt{\xi_{\bm k'}^2+|\Delta(\bm k')|^2 }$. Here  $\Gamma_{\bm k, \bm k'}$ is the chiral Cooper-scattering amplitude, Eq.\eqref{eq:Vkk'_preview}. 
The angle dependence in $\Gamma_{\vec k, \vec k'} $ will lead to an array of different chiral phases.

The general properties of the pairing vertex can be analyzed by suppressing the exchange processes in Eq.\eqref{eq:Vkk'_preview} and focusing on direct 
processes. By a $k$-space gauge transformation, $\Gamma^{\rm dir}$ can be brought to a circular-symmetric form (for details, see Appendix \ref{App: gauge_transform}). 
%
%
We expand $\Gamma^{\rm dir}$ in angular harmonics to obtain:
\begin{equation}
\label{Vm}
    \Gamma^{\rm dir}(\phi) = \sum_m \Gamma_m e^{i m \phi}, \quad \Gamma_m = \int \frac{d\phi}{2\pi} 
    \Gamma^{\rm dir}(\phi) e^{-i m \phi}.
\end{equation}
The values $\Gamma_m$ are readily obtained by inverse Fourier transform, giving a closed-form expression: 
\begin{equation}
\label{Vmdec}
    \Gamma_m = \frac{2\lambda \pi b}{\sinh(2ab)}\,
    e^{
      -\Phi \coth(2ab)
    } I_m\bigg(\frac{\Phi}{ \sinh(2ab)} \bigg) e^{2mab},
\end{equation}
where $I_m$ are modified Bessel functions of the first kind. 
Having the dependence of the quantities  $g_m$ on parameters $b$, $k_F^2$ and $a$, 
we can predict 
chiral phases arising at different electron densities and Berry curvature values.

Reinstating the exchange processes---the antisymmetrization described by Eq. \eqref{eq:Vkk'_preview}---yields dimensionless pairing couplings
\be
g_m=\nu \Gamma_m-(-)^m \nu \Gamma_m
.
\ee
This cancels even-$m$ and doubles odd-$m$ contributions. 
With our convention attractive channels correspond to larger positive $g_m$, with the leading pairing state determined by $g_m$ attaining the maximum value. By plotting $g_m$ for different parameter values, 
we find that leading  odd-$m$ channels 
alternate, resulting in a cascade of transitions between phases with different $m$ (see Fig.~\ref{fig1}). 
Notably, this pattern cannot be explained by the $b$-dependence of the two-particle bound states in Fig.~\ref{fig2_spectrum}, which does not show ground-state switching.


A key insight into this oscillatory behavior can be obtained by noting that the geometric flux through the Fermi sea, Eq.\eqref{eq:Phi_FS},
takes approximately integer values at the phase boundaries. 
Indeed, the boundaries between adjacent phases $m$ and $m+2$ are determined by 
\begin{equation}
\label{bound_eq}
    I_m\bigg({\frac{\Phi}{\sinh(2ab)}}\bigg) = I_{m+2}\bigg({\frac{\Phi}{\sinh(2ab)}}\bigg)e^{4ab}
\end{equation}
Using the asymptotics of the Bessel functions one can show that for the local attraction (large $a$) the boundaries are determined by the equation  $\Phi = \sqrt{(m+1)(m+2)}$, while for the wide-range potential (small $a$) the boundaries are given by small-$a$ asymptotics $\Phi = m+1$. These properties explain the behavior of different phases in the phase diagram shown in Fig.\ref{fig1}b at small and large $ba$. 


We now consider the behavior near a phase boundary where two adjacent allowed
chiral channels, $m$ and $m+2$, become nearly degenerate. For spin-polarized
fermions the allowed pairing channels are odd, $m=1,3,5 \ldots$, so adjacent
phases differ by two units of angular momentum. Near such a boundary it is
sufficient to retain the two dominant components of the gap function,
\[
    \Delta(\phi_{\vec k})
    =
    \Delta_1 e^{im\phi_{\vec k}}
    +
    \Delta_2 e^{i(m+2)\phi_{\vec k}} .
\]
The corresponding mean-field free energy $F(\Delta_1,\Delta_2)$ may be written as
\be 
-T\sum_{\vec k,i\omega_n}
{\rm Tr}\ln
\begin{pmatrix}
    \epsilon_{\vec k}-i\omega_n & \Delta(\phi_{\vec k}) \\
    \Delta^*(\phi_{\vec k}) & -\epsilon_{\vec k}-i\omega_n
\end{pmatrix}
+
\frac{|\Delta_1|^2}{g_m}
+
\frac{|\Delta_2|^2}{g_{m+2}},\quad 
\ee 
After summing over Matsubara frequencies and integrating over momenta normal to
the Fermi surface, one obtains at $T=0$, in the weak-coupling limit,
\begin{equation}\label{F(m,m+2)}
    F
    =
    \nu \oint \frac{d\phi_{\vec k}}{2\pi}
    |\Delta(\phi_{\vec k})|^2
    \ln\frac{|\Delta(\phi_{\vec k})|^2}{E_F^2}
    +
    \frac{|\Delta_1|^2}{g_m}
    +
    \frac{|\Delta_2|^2}{g_{m+2}} .
\end{equation}
This expression, which can be evaluated analytically in closed form~\cite{integral},
makes clear why the transition between neighboring winding sectors is
generically first order. A pure $m$-wave or $(m+2)$-wave state has a uniform
gap magnitude on the Fermi surface, whereas a coherent superposition of the two
components produces an angularly modulated gap magnitude,
\[
    |\Delta(\phi)|^2
    =
    |\Delta_1|^2+|\Delta_2|^2
    +
    2\mathrm{Re}\!\left[
        \Delta_1\Delta_2^* e^{-2i\phi}
    \right].
\]
Such modulation creates gap minima, and in limiting cases nodes, which reduces
the condensation energy relative to a fully chiral single-component state.
Consequently, as the Berry flux $\Phi$ is tuned through the point where
$g_m=g_{m+2}$, the global minimum of the free energy switches discontinuously
from one winding sector to the other. A narrow coexistence regime may occur
close to the boundary, but the dominant behavior is a first-order transition
between two topologically distinct chiral superconducting phases.

The flux interpretation becomes especially transparent in this small-angle
scattering limit: the order parameter may be viewed as a field \(\Delta(\phi_{\vec k})\)
defined along the Fermi surface, and the Berry connection enters as an effective
gauge field in momentum space. The corresponding local Ginzburg--Landau
functional gives a geometric mismatch energy for a winding-\(m\) state,
selecting the chirality whose winding is closest to the Berry flux enclosed by
the Fermi sea. Since only odd \(m\) are allowed in the spin-polarized triplet
problem, tuning \(\Phi\) drives switches between \(m\) and \(m+2\), producing the
quantum-geometric analog of Little--Parks oscillations.


This can be seen explicitly in
the small-$a$ limit of the pairing interaction
$
    V(\vec R_{12}) =\lambda  e^{-a(\vec R_1 - \vec R_2)^2}
$. 
For $a \ll k_F^2$, the interaction is wide-range in $r$-space, which makes the Cooper scattering processes 
$
    (\vec k, -\vec k) \to (\vec k', -\vec k')
$
small-angle, i.e., nearly collinear. These processes, by connecting opposite patches on the Fermi circle, naturally lead to an angle-dependent superconducting order parameter $\Delta(\phi)$ defined on the Fermi circle, with $\phi_{\vec k}$ the azimuthal angle. In this limit, superconductivity is effectively ``patch-local,'' and its behavior is described by a Ginzburg-Landau free energy that is local in $\phi\equiv\phi_{\vec k}$:
\be\label{eq:GL_model}
F(\Delta) = \int d\phi\, \frac{\kappa}2 \Delta^*(\phi)(i\partial_\phi-\Phi)^2\Delta(\phi)
-\alpha|\Delta|^2+\frac{\beta}2|\Delta|^4
,
\ee
where $\Phi$ is $bk_F^2$, Eq.\eqref{eq:Phi_FS}. Minimizing this free energy predicts states with $2\pi m$ phase winding, $\Delta(\phi)\sim e^{im\phi}$, with energies varying with $\Phi$ as
\be\label{eq:LP_oscillations} 
F_m(\Phi)\sim \frac{\kappa}2 
\left( m-\Phi\right)^2|\Delta_0|^2+F_m^{(0)}
.
\ee
The $m$ sectors are stable for $m-1 < \Phi < m+1$. For odd $m$, the phase boundaries at even integer $\Phi$, consistent with the small-$a$ phase diagram in Fig.~\ref{fig1}(b).

A cascade of first-order transitions, which provides a clear experimental signature for this mechanism for chiral superconductivity, has several interesting implications:

One clear manifestation is oscillations of the critical temperature $T_c$, controlled by the Berry flux $\Phi$ through the Fermi sea area and tunable via carrier density. These oscillations are analogous to the Little-Parks effect in conventional superconductors in a magnetic field, where the periodicity of $T_c$ as a function of flux allows one to measure the superconducting flux quantum $hc/2e$. In our case, the oscillations are governed by the 
commensurability between the order-parameter winding and the Berry flux enclosed by the Fermi sea.
This represents a quantum-geometry analog of the Little--Parks effect, providing a striking signature of chiral superconductivity.

Chiral superconducting phases with broken time-reversal symmetry are also expected to carry finite magnetization. Estimating the orbital angular momentum per Cooper pair as $\hbar m$ gives a magnetic-moment density of order $\mu_B m |\Delta|^2$. Its dependence on $m$ means that an external field $H$ reshapes the phase diagram: the energy of a phase with angular momentum $m$ is lowered by $\sim \mu_B m |\Delta|^2 H$, favoring larger $m$ and suppressing smaller $m$. Thus, in a field $H$, each phase boundary in Fig.~\ref{fig1} shifts inward, to smaller $\Phi$, by an amount proportional to $H$.

Last but not least, in a spatially nonuniform system, first-order transitions between phases with different $m$ will generate domains separated by boundaries that host chiral edge modes. The edge supercurrents carried by these modes can be directly probed by spatially resolved scanning magnetometry, while the chiral quasiparticles localized at the domain walls can be detected through signatures in chiral thermal transport.

In summary, this work demonstrates that Berry curvature in electronic bands qualitatively reshapes the superconducting pairing problem, giving rise to a cascade of chiral superconducting states characterized by quantized angular momentum 
$m$. Unlike conventional vortex physics driven by real-space magnetic fields, the Berry-curvature–induced pseudomagnetic field acts in $k$ space, modifying the microscopic structure of Cooper pairs. As a result, superconducting order parameters acquire intrinsic phase winding on the Fermi surface, leading to a family of topologically distinct chiral states, manifested in a number of signatures readily detectable by state-of-the-art experimental techniques.

This work greatly benefited from discussions with 
Patrick Lee, Brian Skinner, Ashvin Vishwanath and Eli Zeldov.

\appendix

\section{
Pairing vertex: 
restoring circular symmetry by a gauge transformation}
\label{App: gauge_transform}

The properties of the pairing vertex $\Gamma_{\vec k, \vec k'} $ can be understood by suppressing exchange processes in Eq.\eqref{eq:Vkk'_preview} and focusing on direct
scattering processes. 
In the center-of-mass frame, Eq.\eqref{eq:H12_CM}, the scattering amplitude for direct processes is $\Gamma^{\rm dir}_{\bm k, \bm k'} = \langle\vec{k}|\lambda e^{-a\hat{R}^2}|\vec{k'} \rangle$. Working in the Landau gauge, we write the quantity $\hat{R}^2$ as
\begin{equation}
 \vec R^2
  =-(\partial_{k_x} +i 2b k_y)^2 - \partial^2_{k_y}.
  \label{eq:R-Landau}
 \end{equation}
 Since $x$ is conserved, we can write  $\Gamma^{\rm dir}_{\vec k, \vec k'}$ as Fourier transform
 \begin{equation}
    \Gamma^{\rm dir}_{\vec k, \vec k'} = \int dx e^{-i (k_x-k_x')x}
    \lambda \tilde{V}_x(k_y,k_y'),
 \end{equation}
 where $\tilde{V}_x(k_y,k_y')$ 
 denotes the matrix element  $ \langle k_y |
 e^{-a\big((x - 2b k_y)^2 - \partial^2_{k_y}\big) }
 | k_y'\rangle$.

  By making a substitution $ 
  k_y \to k_y + \frac{x}{2b}$, we obtain $ 
  \hat R^2
  = 4b^2 k_y^2 - \partial^2_{k_y}$. The matrix element $ 
  \langle k_y | e^{-a (4b^2 k_y^2 - \partial^2_{k_y})}| k_y' \rangle$ can be calculated explicitly via 1D oscillator eigenstates
  \begin{align}
       &\langle k_y | e^{-a (4b^2 k_y^2 - \partial^2_{k_y})}| k_y' \rangle = \sqrt{\frac{4 \pi b}{\sinh(4ab)}} 
      \\ \nonumber
      &\times \exp \bigg[-\frac{b}{\sinh(4ab)}\big((k_y^2+k_y'^2)\cosh(4ab) -2 k_y k_y'\big) \bigg].
  \end{align}
  Substituting back $
  k_y \to k_y - \frac{x}{2b}$ and integrating over $x$, we obtain the final answer
  \begin{align}
  &    \Gamma^{\rm dir}_{\vec k, \vec k'} 
  = \frac{2\lambda \pi b}{\sinh(2ab)}\,
    \exp\!\left[
      -\frac{b}{2} \coth(2ab)\,(\bm k - \bm k')^2
    \right]
    \nonumber \\ &
    \times \exp\!\left[
      -i b (k_y+k_y')(k_x-k_x')
    \right].
    \label{eq:Vkk'_Landau_gauge}
  \end{align}
  Importantly, the form of the vertex function depends on the choice of gauge $\vec a(\vec k)$. 
To transform the result in the Landau gauge, Eq.\eqref{eq:Vkk'_Landau_gauge}, to the one in a symmetric gauge we 
perform a gauge transform for the electron wave functions,
$\bm A \to \bm A + \nabla f$, $\psi \to \psi e^{i f(\vec k)}$. This yields 
\begin{equation}
\Gamma^{\rm dir}_{\vec k, \vec k'}\to \Gamma^{\rm dir}_{\vec k, \vec k'} e^{if(\bm k) - i f(\bm k')}.
\end{equation}
Taking $f(\vec k) =  b k_x k_y$ yields the answer in the symmetric gauge:  
\be 
\vec a(\vec k) -\vec a(-\vec k)=  \vec b \times \vec k,
\ee 
This yields an angular dependence with full cylindrical symmetry:
\begin{equation}
  \Gamma^{\rm dir}_{\vec k, \vec k'}
  = \frac{2\lambda\pi b}{\sinh(2ab)}\,
    \exp \big[
      -\frac{b}{2} \coth(2ab)\,(\bm k - \bm k')^2
     + i b e_{\vec{z}} (\vec k' \times \vec k)\big]
  \label{eq:Vkksolution}
\end{equation}
where $e_{\bm z}(\bm k' \times \bm k)$ denotes a scalar associated with
the vector cross product pointing out of plane. As a sanity check, in the limit $b\to 0$ we recover the vertex function familiar for short-range pairing interaction problem. 

To isolate contributions to pairing with different angular momenta, 
we project 
$\Gamma^{\rm dir}_{\vec k, \vec k'}$ on the angular $m$ harmonic on the Fermi surface. The angle dependence in 
$\Gamma^{\rm dir}_{\vec k, \vec k'}$ is 
\begin{equation}
 \Gamma^{\rm dir}(\phi_{\vec k}-\phi_{\vec k'}) \sim  e^{\Phi\coth(2ab)\cos(\phi_{\vec k}-\phi_{\vec k'})+i \Phi \sin(\phi_{\vec k}-\phi_{\vec k'})},
\end{equation}
where 
$\Phi$ is the geometric flux through the Fermi sea, Eq.\eqref{eq:Phi_FS}. 
Introducing notation
\be 
u =\frac{\Phi}{1-e^{-4ab}},\quad v = \frac{\Phi}{e^{4ab}-1}, 
\ee
we can rewrite the last equation as
\begin{equation}
     \Gamma^{\rm dir}(\phi_{\vec k}-\phi_{\vec k'}) \sim \exp[u e^{i(\phi_{\vec k}-\phi_{\vec k'})} + v e^{-i(\phi_{\vec k}-\phi_{\vec k'})}].
\end{equation}
Expanding $\Gamma^{\rm dir}(\phi_{\vec k}-\phi_{\vec k'})$ in angular harmonics yields  the expression given in the main text, Eqs.\eqref{Vm} and \eqref{Vmdec}. 


\section{Alternative derivation of the pairing vertex} 

The derivation of the scattering amplitude
$V_{\vec k,\vec k'} = \langle \vec k | \lambda e^{-a R^2} | \vec k' \rangle$
given above involves evaluating the matrix element in momentum space in Landau gauge, followed by a transformation to symmetric gauge. In this section, we describe an alternative method based on a direct calculation in symmetric gauge using the exact eigenfunctions and eigenvalues of the operator $\hat{R}^2$:
\begin{subequations}
\begin{equation}
    \psi_{n,m}(\vec k) = c e^{-i m \varphi} e^{-\frac{b k^2}{2}}
    k^{|m|}
    {}_1\! F_1\!\left(-n,\,|m| + 1,\,b k^2\right),
\end{equation}
\begin{equation}
    \rho_{n,m} =  2b(2n + m + |m| +1).
\end{equation}
\end{subequations}
Here the normalization factor
\begin{equation}
    c = b^{\frac{1+|m|}{2}}
    \sqrt{\frac{4 \pi(|m| + n)!}{   n! \, |m|!^2}}
\end{equation}
is defined such that $\int 
d^2 \vec k
|\psi(\vec k)|^2 = 2\pi$. 
This choice ensures the correct normalization of the eigenfunctions in the continuum limit, $b \to 0$. Therefore,
\begin{equation}
    \langle \vec k | \lambda e^{-a R^2} | \vec k' \rangle = \sum_{n,m} \psi_{n,m}^*(\vec k') e^{-a \rho_{n,m}} \psi_{n,m}(\vec k) .
\end{equation}
To proceed, note that the angular dependence in $V_{\bm k, \bm k'}$ is $e^{i m (\phi_{\vec k'}-\phi_{\vec k})}$. Hence, after Fourier transforming,
\begin{equation}
    V_{\bm k, \bm k'} = \sum_m \tilde{V}_m e^{i m (\phi_{\vec k'}-\phi_{\vec k})}
    ,
\end{equation}
we obtain the angular harmonics
\be
\tilde{V}_m = \sum_n \psi_{n,m}^*(k') 
e^{-a \rho_{n,m}}\psi_{n,m}(k)
.
\ee
Thus, the problem is reduced to finding the scattering amplitudes in channels with fixed orbital angular momentum $m$ on the Fermi surface, $|\vec k| = |\vec k'| = k_F$. Substituting the expressions for the eigenfunctions $\psi_{n,m}$, eigenvalues $e^{-a \rho_{n,m}}$, and the Berry flux through the Fermi sea, $\Phi = b k_F^2$, we obtain the following result for $\tilde{V}_m$:
\begin{equation}
    \tilde{V}_m = \frac{4\pi \lambda b}{|m|!^2} e^{-\Phi}
    (\Phi)^{|m|} e^{-2ab  \left( |m| + m + 1
  \right)}\sum_n A_{m,n},
\end{equation}
where
\begin{equation}
    A_{m,n} = \frac{(|m| + n)!}{  n! \,}
    {}_1\! F_1\!\left(-n,\,|m| + 1,\,\Phi\right)^2 e^{-4 b a n}.
\end{equation}

The hypergeometric functions in the sum can be expressed in terms of generalized Laguerre polynomials:
\begin{equation}
    {}_1\! F_1\!\left(-n,\,|m| + 1,\,\Phi\right) = \frac{n!|m|!}{(|m|+n)!} L^{|m|}_n (\Phi ).
\end{equation}

We can now use the well-known Hille--Hardy formula to evaluate the sum:
\begin{multline}
    \sum_{n=0}^{\infty} \frac{n!}{(n+a)!} L^a_n(x) L^a_n(y) t^n = \mathcal{B}_a(x,y,t),
    \\
    \mathcal{B}_a(x,y,t) = \frac{ (xyt)^{-a/2} }{1-t} e^{-\frac{t(x+y)}{1-t}}I_a\bigg(\frac{2\sqrt{xyt}}{1-t}\bigg).
\end{multline}
This gives the following expression for the sum $\sum_n A_{n,m}$:
\begin{multline}
    \sum_n A_{n,m} =  \frac{(|m|!)^2 (\Phi )^{-|m|}  e^{2ab|m|}}{1- e^{-4 ab}}
    \exp\bigg[-\frac{2\Phi e^{-4 ab}}{1-e^{-4ab}}\bigg] 
    \\
    \times   I_{|m|} \bigg(\frac{2\Phi e^{-2ab}}{1-e^{-4ab}}\bigg).
\end{multline}
Finally, we obtain
\begin{equation}
    \tilde{V}_m = \frac{2\lambda \pi b}{\sinh(2ab)}\,
    e^{
      -\Phi \coth(2ab)
    } I_{|m|}\bigg(\frac{\Phi}{ \sinh(2ab)} \bigg) e^{-2mab}.
\end{equation}
These Fourier coefficients $\tilde{V}_m$ are related to the coefficients $V_m$ defined in the main text. The difference is that the angular dependence in the Fourier series for $V_{\vec k,\vec k'}$ is parameterized here by the functions $e^{i m (\phi_{\vec k'}- \phi_{\vec k})}$, whereas in the main text we use
$V_{\vec k,\vec k'} = \sum_m V_m e ^{i m (\phi_{\vec k}- \phi_{\vec k'})}$. Thus,
\begin{equation}
    V_{\vec k,\vec k'} = \sum_m V_m e ^{i m (\phi_{\vec k}- \phi_{\vec k'})} = \sum_m \tilde{V}_m e ^{i m (\phi_{\vec k'}- \phi_{\vec k})}.
\end{equation}
This immediately gives the coefficients $V_m$:
\begin{equation}
    V_m = \frac{2\lambda \pi b}{\sinh(2ab)}\,
    e^{
      -\Phi \coth(2ab)
    } I_{|m|}\bigg(\frac{\Phi}{ \sinh(2ab)} \bigg) e^{2mab},
\end{equation}
which is exactly the result derived in the main text.


\begin{thebibliography}{99}
\bibitem{integral}
Integration over the angle $\phi_{\vec k}$ can be carried out using the identity
\begin{align}
& \int_{0}^{2\pi} \bigl|a + b e^{i\phi}\bigr|^2
\ln\!\bigl(\bigl|a + b e^{i\phi}\bigr|^2\bigr)\,\frac{ d\phi}{2\pi}
=
\\ \nonumber
& \bigl(|a|^2+|b|^2\bigr)  \left(1+\ln\max(|a|^2,|b|^2)\right)
-\bigl||a|^2-|b|^2\bigr|.
\end{align}
\end{thebibliography}
\end{document}